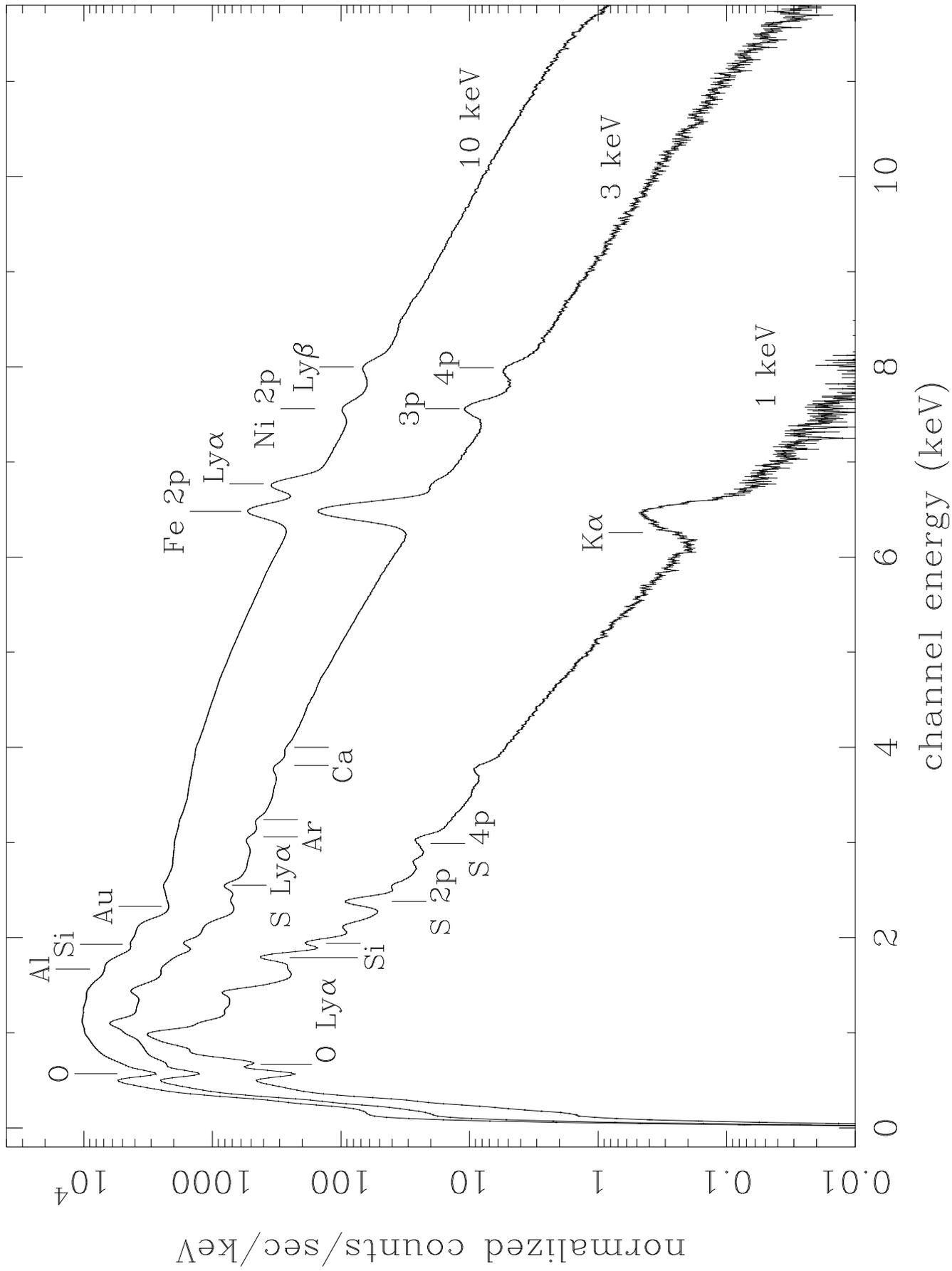

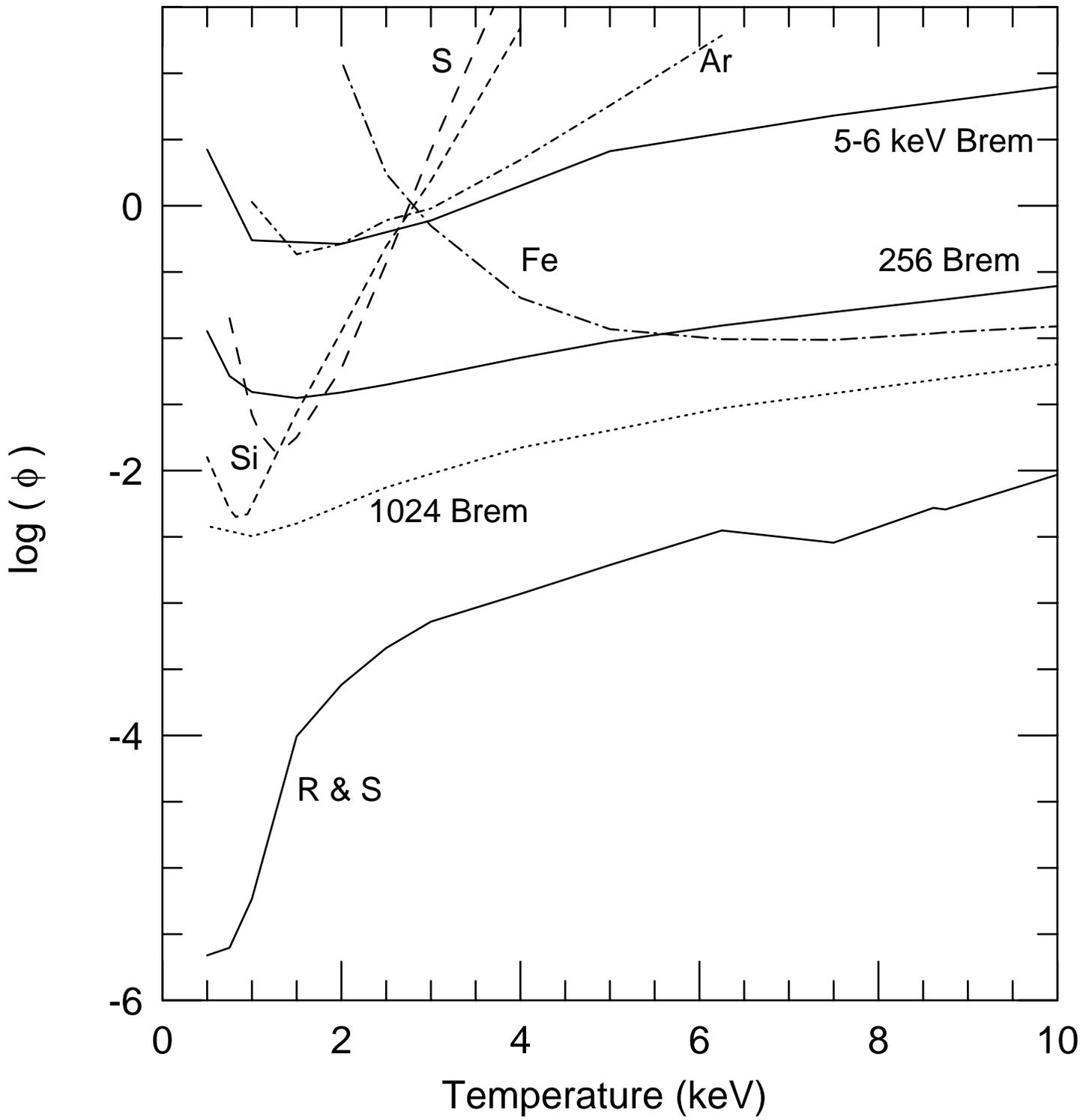

# Statistical Uncertainties in Temperature Diagnostics for Hot Coronal Plasma Using the ASCA SIS

Douglas A. Swartz[1,2] S. L. O'Dell[1], M. E. Sulkanen[1] & A. F. Tennant, Jr.[1]

## ABSTRACT

Statistical uncertainties in determining the temperatures of hot (0.5 keV to 10 keV) coronal plasmas are investigated. The statistical precision of various spectral temperature diagnostics is established by analyzing synthetic ASCA Solid-state Imaging Spectrometer (SIS) CCD spectra. The diagnostics considered are the ratio of hydrogen-like to helium-like line complexes of $Z \geq 14$ elements, line-free portions of the continuum, and the entire spectrum. While fits to the entire spectrum yield the highest statistical precision, it is argued that fits to the line-free continuum are less susceptible to atomic data uncertainties but lead to a modest increase in statistical uncertainty over full spectral fits. Temperatures deduced from line ratios can have similar accuracy but only over a narrow range of temperatures. Convenient estimates of statistical accuracies for the various temperature diagnostics are provided which may be used in planning ASCA SIS observations.

*Subject headings:* Methods: Data Analysis – X-Rays: General – Galaxies: Clustering

## 1. Introduction

As x-ray telescope and detector technology improves, a greater number and diversity of spectral features become accessible for plasma temperature diagnostics. Historically, the use of specific x-ray spectral plasma diagnostics has been limited by poor energy resolution, narrow passbands, and small numbers of detected photons. Most x-ray emission line sources are thermal plasmas which also emit a strong thermal bremsstrahlung component. Thus, any program to identify sensitive temperature diagnostics while making the most efficient use of the available photons must address continuum as well as line diagnostic features. In this work, the statistical precision of a variety of plasma temperature diagnostics are compared. If systematic errors are not dominant, then identifying the most precise x-ray temperature diagnostic for a particular photon-limited observation reduces to determining the statistical uncertainties associated with the corresponding spectral features.

---

[1]Astrophysics Branch, ES-84, NASA-Marshall Space Flight Center, Huntsville, AL 35812

[2]NAS/NRC Resident Research Associate

The number and classes of temperature diagnostics that can be considered is limited by the capabilities of the x-ray detector under study. The moderate spectral resolution and broad-band sensitivity of the Advanced Satellite for Cosmology and Astrophysics (ASCA) mission's Solid-state Imaging Spectrometer (SIS) CCDs are the first to provide access to numerous spectral features between $\sim 0.5$ and $\sim 10$ keV. It is now pertinent to determine which of the many spectral temperature diagnostics is the most statistically precise. In this work, the ratios of the H-like to the He-like line complexes of $Z \geq 14$ elements, line-free portions of the continuum, and the entire spectrum are investigated as potential temperature diagnostics by estimating the statistical uncertainties in the temperature measurements based upon synthetic ASCA SIS spectra of coronal-equilibrium x-ray sources. Temperature-dependent figure of merit functions are defined for each diagnostic that allow convenient, quantitative estimates of the statistical uncertainties to be made from estimates of the source flux and the exposure time.

## 2. A Figure of Merit

Only statistical uncertainties are considered in this work. Systematic effects are minimized or eliminated by using the same set of physical assumptions, atomic data, and instrument/telescope response functions in the simulation as used in the analysis. The statistical uncertainty, $\sigma_T$, in the derived temperature is proportional to the square root of the number of photons within the relevant spectral features. This number of photons is the product of the integration time, the integrated flux, the effective area of the detector, and the detector efficiency. Defining $F_{-11}$ to be the flux in the 2-10 keV band in units of $10^{-11}$ erg cm$^{-2}$ s$^{-1}$ and $t_4$ to be the exposure time in units of $10^4$ s, the uncertainty in the temperature can be written in terms of these fiducial quantities as

$$\left(\frac{\sigma_T}{T}\right)^2 = \frac{\phi(T)}{F_{-11} t_4} \quad (1)$$

where the uncertainty is associated with some pre-chosen confidence level. The variance coefficient, $\phi(T)$, is unique to each diagnostic. It includes characteristic detector efficiency and effective area dependencies. The temperature dependence of $\phi(T)$ comes primarily from the fact that the flux in the diagnostic feature is a temperature-dependent function of the flux in the 2-10 keV band. The quantity $\phi(T)$ is a figure-of-merit function for the temperature diagnostic: It measures the statistical precision to which the temperature can be determined from a given observation. A lower value of $\phi(T)$ reflects a higher statistical precision.

The quantities, $\phi(T)$, for ASCA SIS observations were derived by analyzing synthetic spectra of isothermal coronal plasma ($0.5 < T < 10$ keV) within the XSPEC data-analysis package (v8.4; Shafer et al. 1991). The spectra were generated from 'Raymond-Smith' (hereafter R-S) coronal equilibrium models (based upon the original work of Raymond & Smith 1977) and convolved with a combined telescope and detector response function. Poisson counting statistics were applied. The four quantities parameterizing the spectra are the plasma temperature, metallicity, source redshift, and source emission integral. The metallicity (one-half cosmic abundances) and redshift ($z = 0.03$) chosen represent typical values for intracluster gas in clusters of galaxies (Arnaud et al. 1992). The metallicity sets the strengths of emission lines relative to the H- and He-dominated bremsstrahlung continuum, while the redshift positions the spectral lines relative to any structure within the detector energy efficiency curve. The emission integral and the integration time for the spectra were chosen concurrently to produce a high fluence and large numbers of detected photons in order to minimize random errors in the analysis. The spatial extent of the source, x-ray background, and interstellar absorption are neglected in the construction and analysis of the spectra.

From the numerous available ASCA response functions, one with 256 channels (designed for early ASCA observation feasibility studies) and another with 1024 channels (based on in-orbit calibration) are used in this work to illustrate the effect of energy resolution on the utility of the various temperature diagnostics. Calibration errors are eliminated by using the same representation of the combined instrument and telescope response function in both generating and analyzing each spectrum.

## 3. Temperature Diagnostics

Temperature diagnostics were selected for this study based upon the expected strength of the spectral features in the temperature range 0.5 to 10 keV, on their temperature sensitivity, and on the ability of the spectral components to be resolved by ASCA SIS at its highest resolution. Within these criteria, several classes of temperature diagnostics can be identified (Figure 1): (1) the H-like to He-like line ratios of $Z \geq 14$ metals ($n = 2$ to $n = 1$ transitions), (2) the line-free portions of the continuum, and (3) the entire spectrum.

### 3.1. Line Ratios

Ratios of lines from the same ionization stage are either unresolved or strongly blended with other lines in ASCA observations, but the H/He-like line ratio can be resolved even with the 256 channel response function (although nearby lines of other elements sometimes blend with the He-like or H-like complexes at certain temperatures – such as with the Ar He-like complex as discussed in § 5). In the R-S model, the He-like complex arising from the $n = 2$ level includes the $^1P$ resonance, $^3P$ intercombination, $^3S$ forbidden, and dielectronic satellite lines (the latter are modeled as a single line within the $n = 2$ complex). The H-like feature contains the single Ly-$\alpha$ line. (No H-like dielectronic satellites are included.) Spectral lines are modeled within XSPEC by fitting a selected portion of the continuum containing the lines of interest and excluding the remaining energy channels from the data. Lines are modeled with gaussian profiles and, concurrently, the underlying continuum by a bremsstrahlung component. The line fitting is optimized by choosing a broad bandwidth but with a minimum number of spectral lines. This minimizes the uncertainty in the continuum parameters thereby reducing the uncertainty in the line strengths. The bandwidths chosen are 1.73-2.21 keV for Si, 2.09-2.90 keV for S, 2.71-3.40 keV for Ar, and 5.29-7.36 keV for Fe. Line ratios are determined by the number of photons in the gaussian components and the corresponding temperature is derived by interpolating these ratios along tabulated values of the ratio as a function of temperature. The tabulated values were calculated using the Raymond and Smith code (outside of XSPEC) and assume the same physics and atomic data as was used to construct the synthetic spectra.

### 3.2. Continua

To a very good approximation, the continuum at temperatures $T \geq 0.5$ keV arises solely from H and He bremsstrahlung. The temperature is, therefore, determined by the shape or slope of the continuum so that the precision of the derived temperature is improved by increasing the bandwidth sampled. The continuum is modeled within XSPEC by first excluding the obvious line features from the data and then fitting the remaining channels with a bremsstrahlung spectrum. Weak lines can be identified in the fit residuals, excluded, and a new fit made to the data in the remaining channels. Using high fluence synthetic spectra ensures lines can be easily identified. This selection of the bandwidth is performed for the spectra at 2 keV (5 keV for fitting only the 5.25-to-6.3 keV band, see below) and, as for lines, the bandwidth is not altered for succeeding spectra in order to properly evaluate $\phi(T)$ even though fewer lines appear at the higher temperatures and the bandwidth could be increased in practice. The free-free Gaunt factors used in XSPEC for the bremsstrahlung model are the same as those used within the Raymond and Smith code. In addition to the entire line-free spectrum, a contiguous line-free portion spanning $\sim 5.25$ to $\sim 6.3$ keV is also analyzed and a figure of merit function derived.

### 3.3. Full Spectrum

Models designed to fit the entire spectrum maximize the usage of the detected photons thereby minimizing the statistical errors in the result. Systematic errors are minimized in our analysis by fitting R-S models to the synthetic spectra. The resulting values of $\phi(T)$ represent the ideal minimum values attainable from actual observations.

### 4. Results

The coefficient, $\phi(T)$, defined at the $1\sigma$ confidence level, is shown in Figure 2 for the H/He-like line ratios of Si, S, Ar, and Fe (Ca and Ni cosmic abundances are too low to be considered). Also shown are the results from fits to all line-free spectral bands using two different energy resolution response functions, fits to the contiguous 5.25-to-6.3 keV continuum band, and R-S fits to the entire spectrum. The figure-of-merit functions can be used to estimate the uncertainties expected from ASCA SIS observations. For example, from Equation 1 and Figure 2, to determine the temperature of a source with $F_{-11} \sim 1$ to within 10% at the $1\sigma$ level by fitting to the full continuum spectrum requires an exposure time $t \sim 63$ ks if the source temperature is 10 keV using the higher-resolution response function.

To derive $\phi(T)$, the $\chi^2$ fit statistic was minimized to obtain the best fit parameter values and the $1\sigma$ errors for each interesting parameter were determined within the XSPEC data analysis program. The temperature is not a model parameter for line ratio diagnostics. Instead, the statistical uncertainties in the line ratios are computed from the uncertainties in the individual line strengths by error propagation (assuming the errors are uncorrelated). This defines a range of line ratios that maps to a (tabulated) range of temperatures to define the statistical uncertainties in the derived temperature.

The R-S model fit to the entire spectrum produces the smallest statistical uncertainties and hence the lowest $\phi(T)$ curve in Figure 2. The same result was found by fitting either the low- or the high-resolution spectra. The small kink in this curve at $\sim 7$ keV is a consequence of the interpolation method of fitting tabulated R-S models within XSPEC. The importance of line and bound-free emission features in spectra at temperatures $T \lesssim 3$ keV is clearly evident in the R-S fits. These features are more temperature-sensitive than is the bremsstrahlung component, thus the uncertainty in the fitted temperatures is greatly decreased at low temperatures. This tendency holds true even for the low resolution spectra studied here although it is not expected to hold for arbitrarily low resolution.

At temperatures above $\sim 3$ keV, continuum photons dominate the spectrum. The uncertainty in the temperature derived from the R-S fit is, therefore, dominated by the uncertainty in fitting the continuum at these high temperatures. This accounts for the similarity between the temperature dependence of $\phi(T)$ for the continuum bands and for the R-S model fits at temperatures above $\sim 3$ keV. Comparison of the line-free continua results reveals a lower value of $\phi(T)$ at higher spectral resolution. This occurs because a smaller fraction of the data channels contain lines in the higher resolution spectra. Thus, the two curves representing fits including all continuum channels differ only by the number of photons retained in the fitted spectrum for $T \gtrsim 2$ keV.

Fits to the 5.25-to-6.3 keV continuum band were performed to illustrate an additional uncertainty entering the temperature determined from fitting narrow continuum bands. The figure of merit for the 5.25-to-6.3 keV band does not simply scale as the number of photons as did the full continuum fits. The temperature parameter of the bremsstrahlung model defines the shape of the model spectrum, but the slope of the curve through the data becomes indeterminate in the limit of negligible bandwidth. Thus the value of $\phi(T)$ for the narrow 5.25-to-6.3 keV continuum band exceeds that for the entire continuum by more than merely the ratio of the number of photons in the bands (equal to $\sim 0.6$ at 6 keV).

The figure of merit functions for the H/He-like line ratios for each element reaches a minimum at some characteristic temperature and increases for both higher and lower temperatures. This is most readily seen in the curves for Si and S which have minima at $\sim 0.8$ and $\sim 1.2$ keV, respectively. This is an effect of the changing ionization state of the plasma with temperature: The He-like lines decrease in strength relative to the continuum as the temperature rises and the strength of the H-like line declines as the temperature falls. The line strength uncertainty increases as the line becomes weak relative to the continuum producing the characteristic shape of the $\phi(T)$ curve. Iron is predominantly in the He-like and H-like ionization states over a broad range of temperatures according to the R-S model ionization balance. Combined with the high relative abundance of Fe, the H/He-like Fe line ratio is a superior temperature diagnostic than the H/He-like line ratio for other elements for spectral temperatures as low as ∼ 3 keV. Line strengths depend linearly on the (ionic) abundances. The values of $\phi(T)$ for the line ratios can therefore be scaled accordingly for different source metallicities (though, in practice, the metallicity may not be known).

## 5. Additional Uncertainties

Only statistical uncertainties have been considered in determining $\phi(T)$. Systematic errors introduce a deviation of the derived temperature from the actual source temperature. The ratio of the derived temperature to the actual source temperature is shown against the source temperature for each diagnostic in Figure 3. By eliminating other sources of systematic error and by minimizing statistical errors by choosing spectra with high count rates, the inaccuracies indicated in Figure 3 simply reflect the inappropriateness of the models used to fit the data: Bound-free contributions were neglected in the (bremsstrahlung) continuum model. These contributions are negligible except for $T \lesssim 2$ keV where the relatively flat bound-free spectrum causes temperature overestimates. For H/He-like line ratios, particularly S and Ar, modeling line blends as single (Gaussian) features can be a poor model. Si Ly-$\beta$ lies near the S He-like $2p$ complex. By not accounting for this Si line in modeling the H/He-like S ratio, the temperature is underestimated at high temperatures where the Si Ly-$\beta$ line becomes strong. Similarly, the S Ly-$\beta$ line is blended with the Ar He-like complex. Though this line was included as a separate line in the model, the uncertainty in the He-like Ar line is large because of this blending and the derived temperature is inaccurate.

In practical applications, additional systematic uncertainties arise which were eliminated in this work. These include uncertainties in the atomic data and ionization equilibria which primarily affect the H/He-like line ratios and calibration errors which are important to the broadband diagnostics. Modeling the entire spectrum, including lines, gives the highest statistical precision; but it is subject to systematic errors resulting from uncertainties in the atomic data and atomic models, the ionization state of the gas, calibration, and elemental abundances. Foreground absorption, either galactic in origin or intrinsic to the source, is another source of uncertainty in fits to broad spectral bands. The effect is insignificant for low column densities; but, for $N_H \sim 10^{23}$ atoms cm$^{-2}$, $\phi(T)$ is increased by a factor of $\lesssim 2$, and the best-fit temperature is systematically increased, especially for hotter plasmas.

## 6. Conclusions

The statistical uncertainties in determining the temperatures of hot coronal plasmas have been investigated by analyzing synthetic ASCA CCD spectra through standard data analysis procedures. Fits to the entire spectrum give the highest statistical precision. Fits to line-free spectral bands provide lower but comparable precision – especially at temperatures $T\gtrsim 3$ keV where the spectrum is dominated by bremsstrahlung emission. Ratios of line strengths are also valuable temperature diagnostics but over restricted temperature ranges. The only temperature-sensitive line ratios identified as accessible to ASCA spectroscopy are composed of the H-like and He-like transitions.

A figure-of-merit function was designed to illustrate the relative statistical uncertainties associated with each temperature diagnostic. These functions can be used to estimate statistical uncertainties in actual ASCA SIS observations.

The appropriateness of the underlying physical models used to fit the data were qualitatively addressed. The accuracy of bremsstrahlung continuum fits are better than $\sim 5\%$ except for $T\lesssim 2$ keV where bound-free contributions arise. Single Gaussian profile fits to complex line blends may be less accurate, but for the H-like and He-like line blends modeled here (based on R-S atomic models) they are usually accurate to $\sim 10\%$ over appropirate temperature ranges. Other sources of systematic error may arise in practical applications such as calibration and foreground absorption uncertainties (which mainly affect broadband diagnostics) and atomic data, ionization equilibria, and elemental abundance uncertainties.

---

Fig. 1.— Synthetic spectra of isothermal coronal equilibrium plasmas are shown for three plasma temperatures (1, 3, and 10 keV). All spectra have been convolved with a 1024-channel response function appropriate to an ASCA SIS (CCD) and telescope combination. The O, Al, Si, and Au labels at upper left indicate instrumental CCD and telescope edges. Labeled emission lines include H-like (Ly-$\alpha$ and Ly-$\beta$) and He-like ($np$) complexes. All those above 5 keV arise from Fe except for the Ni $2p$ line at 7.56 keV (which is blended with Fe $3p$). Line complexes labeled Si, Ar, and Ca in the two cooler spectra indicate both He-like (at lower energy) and H-like line groups. Fe and Ni L-shell lines produce the broad blend extending from $\sim 0.7$ to $\sim 1.5$ keV in the 3 keV and 1 keV spectra.

Fig. 2.— The figure of merit functions for several temperature diagnostics are shown against the plasma temperature. Labels denote the ratios of H-like to He-like line complexes in iron (*Fe*), argon (*Ar*), sulfur (*S*), and silicon (*Si*); fits to the line-free portions of the continuum using a 256 channel response function (*256 Brem*) and a 1024 channel response function (*1024 Brem*); fits to the line-free continuum spanning 5.25 to 6.3 keV (*5-6 keV Brem*); and fits to the entire spectrum using a Raymond-Smith model (*R&S*).

Fig. 3.— The ratio of the derived temperature to the actual temperature is shown against the actual temperature for the spectral diagnostics shown in Figure 2. The Raymond-Smith fit to the entire spectrum is not shown as it deviates imperceptibly from $T/T_o = 1$. Shown are the results for the H/He-like line ratios in Fe (*dot-long dashed line*), Ar (*dot-short dashed line*), S (*long dashed line*), and Si (*short dashed lines*); fits to the 5.2-6.4 keV band and to the line-free portions of the continuum using the 256 channel response matrix (*solid lines*); and fits to the line-free portions of the continuum using the 1024 channel response matrix (*dotted line*).